\documentclass[prd, twocolumn, superscriptaddress, nofootinbib, notitlepage, floatfix]{revtex4-1}
\usepackage{xcolor}
\usepackage{graphicx}
\usepackage{wrapfig}
\usepackage{amsmath}
\usepackage{amsfonts}
\usepackage{multirow}
\usepackage{slashed}
 \usepackage{physics}
 \usepackage[colorlinks=true, linkcolor=blue, citecolor=blue, urlcolor=blue]{hyperref}
\begin{document}
\title{Bethe-Salpeter amplitudes of Upsilons}
\author{Rasmus Larsen}
\affiliation{Physics Department, Brookhaven National Laboratory, Upton, New York 11973, USA.}
\email{rlarsen@bnl.gov}
\author{Stefan Meinel}
\affiliation{Department of Physics, University of Arizona, Tucson, Arizona, USA.}
\affiliation{RIKEN-BNL Research Center, Brookhaven National Laboratory, Upton, New York 11973, USA.}
\author{Swagato Mukherjee}
\affiliation{Physics Department, Brookhaven National Laboratory, Upton, New York 11973, USA.}
\author{Peter Petreczky}
\affiliation{Physics Department, Brookhaven National Laboratory, Upton, New York 11973, USA.}
\begin{abstract}
  Based on lattice non-relativistic QCD (NRQCD) studies, we present results for
  Bethe-Salpeter amplitudes for \(\Upsilon(1S)\), \(\Upsilon(2S)\), and
  \(\Upsilon(3S)\) in vacuum as well as in quark-gluon plasma. Our study is based on
  \(2+1\) flavor $48^3 \times 12$ lattices generated using the Highly Improved
  Staggered Quark (HISQ) action and with a pion mass of $161$ MeV. At zero
  temperature the Bethe-Salpeter amplitudes follow the expectations based on
  non-relativistic potential models. At non-zero temperatures, the interpretation
  of Bethe-Salpeter amplitudes turns out to be more nuanced but consistent with
  our previous lattice QCD study of excited Upsilons in quark-gluon plasma.
\end{abstract}
\date{\today}
\maketitle
\section{Introduction}

Potential models give a good description of the quarkonium spectrum below the open charm
and bottom thresholds; see e.g., Refs. \cite{Brambilla:2004wf,Brambilla:2010cs} for reviews.
Even some of the states above the threshold are also reproduced well within this model.
Potential models can be justified using an effective field theory approach
\cite{Brambilla:1999xf,Brambilla:2004jw}. This approach is based on the idea
that for a heavy quark with mass $m$, there is a separation of energy scales related to
the quark mass, inverse size of the bound state, and binding energy, $m \gg mv \gg m v^2$,
with $v$ being the velocity of the heavy quark inside the quarkonium bound state.
The effective field theory at scale $m v$ is the non-relativistic QCD (NRQCD), where the heavy quark
and anti-quark are described by non-relativistic Pauli spinors
and pair creation is not allowed in this theory \cite{Caswell:1985ui}.
The effective
theory at scale $mv^2$ is potential NRQCD (pNRQCD), and the quark anti-quark potential appears as
a parameter of the pNRQCD Lagrangian. Potential model appears as the tree level approximation of
pNRQCD \cite{Brambilla:2004jw}. Non-potential effects are manifest in the higher order corrections.
For very large quark mass, $v \sim \alpha_s \ll 1$. Therefore,
the large energy scales can be integrated out perturbatively \cite{Brambilla:1999xf,Brambilla:2004jw}.
However, for most of the quarkonium states realized in nature this condition is not fulfilled.
If $\Lambda_{QCD} \gg m v^2$,
all the energy scales can be integrated out non-perturvatively and the potential is given in terms of
Wilson loops calculated on the lattice \cite{Brambilla:1999xf,Brambilla:2004jw}.
So, in this limit, too, the potential description is justified.
However, for many quarkonia,
$\Lambda_{QCD} \simeq m v^2$, and it is not clear how to justify the potential models.

In potential models, one can also calculate the quarkonium wave function. On the
other hand, in lattice QCD we can calculate the Bethe-Salpeter amplitude, which in the
non-relativistic limit reduces to the wave function. Thus, one can use the Bethe-Salpeter
amplitude for further tests of the potential models. In particular, one can also reconstruct
the potential from the Bethe-Salpeter amplitude \cite{Ikeda:2011bs,Kawanai:2011xb,Kawanai:2011jt,Kawanai:2013aca,Nochi:2016wqg}.
Most of these studies focused on quark masses close to or below the charm quark
mass, though in Ref. \cite{Kawanai:2013aca} quark masses around the bottom quark have also been
considered. The resulting potential turned out to be similar to the static potential calculated on
the lattice, but some differences have been found. The potential description is expected to work better
for larger quark masses, and therefore the bottomonium is best suited for testing this approach.
Studying the bottomonium on the lattice using a fully relativistic action is more difficult
because of the large cutoff effects and the rapid fall-off of the correlators.
One of our aims is to test the potential model by calculating the bottomonium Bethe-Salpeter amplitude
using lattice NRQCD \cite{Lepage:1992tx,Thacker:1990bm},
which is very well suited for studying the bottomonium \cite{Davies:1994mp,Meinel:2009rd,Meinel:2010pv,Hammant:2011bt,Dowdall:2011wh,Daldrop:2011aa,Lewis:2012ir,Wurtz:2015mqa}.

The existence and the properties of quarkonia in the hot medium attracted a lot of
attention in the last 30 years. It was proposed a long time ago that quarkonium
production in heavy-ion collisions can be used to probe quark-gluon plasma (QGP)
formation \cite{Matsui:1986dk}. The study of in-medium properties of quarkonia and
their production in heavy ion collisions is an extensive research program; see e.g.,
Refs. \cite{Aarts:2016hap,Mocsy:2013syh,Bazavov:2009us} for reviews. The
in-medium properties of quarkonia as well as their dissolution (melting) are encoded
in the finite temperature spectral functions. Quarkonium states show up as peaks in
the spectral function that become broader as the temperature increases and eventually
disappear above some temperature (\(T\)). The temperature above which no peaks in the
spectral function can be identified is often called the melting temperature.
Reconstructing quarkonium spectral functions from lattice calculations at non-zero
temperature appeared to be very challenging (see, e.g., discussions in
Refs.~\cite{Wetzorke:2001dk,Datta:2003ww,Jakovac:2006sf,Mocsy:2007yj}). The study of
Bethe-Salpeter amplitudes has been proposed as an alternative method to address this
problem. The idea behind this approach is to compare the Bethe-Salpeter amplitude
calculated on the lattice with the expectations of the free field theory that would
indicate an unbound heavy quark anti-quark pair. Bethe-Salpeter amplitudes at
non-zero temperature for charmonium have been calculated in previous lattice QCD
studies~\cite{Umeda:2000ym,Ohno:2008cc,Umeda:2008kz,Ohno:2011zc,Evans:2013yva,Allton:2015ora},
but presently our understanding regarding the interpretations of quarkonia
Bethe-Salpeter amplitudes at \(T>0\) remains murky. Although using a weak-coupling
approach it is possible to generalize the potential description to non-zero
temperature~\cite{Laine:2006ns,Brambilla:2008cx}, it is unclear if such an
approach and the interpretations of quarkonia Bethe-Salpeter amplitudes are applicable
in the temperature regime of interest. In this paper, we focus on lattice NRQCD based
determinations of Bethe-Salpeter amplitudes of \(\Upsilon(1S)\),  \(\Upsilon(2S)\),
and \(\Upsilon(3S)\) states at \(T>0\). By comparing with the corresponding \(T=0\)
results, where the  interpretations of Bethe-Salpeter amplitudes are more
straightforward, we point out and discuss subtleties associated with interpretations
of Bethe-Salpeter amplitudes  at \(T>0\).

\section{Bethe-Salpeter amplitudes at \(\mathbf{T=0}\)}

To define the Bethe-Salpeter amplitude for the bottomonium we consider the correlation function
\begin{equation}
\tilde C^r_{\alpha}(\tau) = \expval{O^r_{qq}(\tau) \tilde O_{\alpha}(0)},
\label{def_Calpha}
\end{equation}
where $\tilde O_{\alpha}$ is the meson operator that has a good overlap with a
given quarkonium state $\alpha$ and $O^r_{qq}$ is a point-split meson operator
with the quark and antiquark fields separated by distance $r$,
\begin{equation}
O^r_{qq}(\tau)= \sum_{\mathbf{x}} \bar q(\mathbf{x},\tau) \Gamma q(\mathbf{x}+\mathbf{r},\tau).
\end{equation}
Here, $\Gamma$ fixes the quantum number of the meson. Furthermore, in the present work we use Coulomb gauge fixed ensembles to define the  expectation value.
Inserting a complete set of states we obtain the following spectral decomposition of the correlator:
\begin{equation}
\tilde C_{\alpha}^r(\tau)=\sum_n \mel{0}{O^r_{qq}(0)}{n} \mel{n}{\tilde O_{\alpha}(0)}{0}e^{-E_n \tau}.
\end{equation}
Assuming that only one state \(\ket{\alpha}\) contributes at large $\tau$, which is correct for an appropriately chosen
$\tilde O_{\alpha}$, at large Euclidean time we have
\begin{equation}
\tilde C^r_{\alpha}(\tau)=  A_{\alpha}^{*} \mel{0}{O^r_{qq}(0)}{\alpha} e^{-E_{\alpha} \tau},
\end{equation}
where $A_{\alpha}^*=\mel{\alpha}{\tilde O_{\alpha}(0)}{0}$.
The matrix element
\begin{equation}
\phi_{\alpha}(r) = \mel{0}{O^r_{qq}(0)}{\alpha}
\end{equation}
is called the Bethe-Salpeter (BS) amplitude and describes
the overlap of the quarkonium state $|\alpha\rangle$
with the state that is obtained by letting the two field operators at distance $r$ act on the vacuum.
In the non-relativistic limit, it reduces to the wave function of
the given quarkonium state. Thus, up to normalization factor, the Bethe-Salpeter amplitude
is given by the large $\tau$ behavior of $\exp(E_{\alpha} \tau) C_{\alpha}(\tau)$, with $E_{\alpha}$ being
the energy of quarkonium state $|\alpha\rangle$, which is also calculated on the lattice.
In the following, we will use the terms BS amplitude and wave function interchangeably.

As mentioned in the Introduction, we aim to calculate the bottomonium BS amplitudes
using NRQCD. We performed calculations using 2+1 flavor gauge configurations generated
by HotQCD with the highly improved staggered quark (HISQ) action \cite{Bazavov:2011nk,Bazavov:2014pvz}.
The strange quark mass was fixed to its physical value, while the light quark masses correspond
to the pion mass of $161$ MeV in the continuum limit \cite{Bazavov:2011nk,Bazavov:2014pvz}.
We use the same NRQCD formulation as in our previous study \cite{Larsen:2019bwy,Larsen:2019zqv}.
For the calculations at zero temperature, we use $48^4$ lattices and $\beta=10/g_0^2=6.74$ corresponding
to lattice spacing $a=0.1088$ fm. We use $192$ gauge configurations in our analysis with eight sources
per configuration.

To construct the meson operators that have the optimal projection we start with the source~\cite{Larsen:2019zqv}
\begin{equation}
O_i(\tau,\mathbf{x})=\sum_\mathbf{r} \psi_i(\mathbf{r}) \bar q(\tau,\mathbf{x}) \Gamma q(\tau,\mathbf{x}+\mathbf{r}).
\label{Oi_def}
\end{equation}
Here, $\psi_i(r)$ is the trial wave function of the $i$th bottomonium state obtained by solving
the Schr\"odinger equation with the Cornell potential modified by discretization effects~\cite{Meinel:2010pv}.
Since $G_{ij}(\tau)=\expval{O_i(\tau)O_j(0)}$ is non-zero (though small) also for $i\neq j$, we have to solve the generalized
eigenvalue problem
\begin{equation}
G_{ij}(\tau) \Omega_{j\alpha}=\lambda_{\alpha}(\tau,\tau_0) G_{ij}(\tau_0) \Omega_{j\alpha}
\label{Omega_def}
\end{equation}
to obtain the optimized operator for bottomonium state $\alpha$,
\begin{equation}
\tilde O_{\alpha}=\sum_j \Omega_{j \alpha} O_j.
\end{equation}
The value of $\tau_0$ is arbitrary to some extent but should be considerably smaller than $\tau$. 
Choosing larger $\tau_0$ helps suppressing higher lying states, i.e.,
states with energies larger that the energy of $\Upsilon(3S)$. However,
the operators $O_i$ in Eq. (\ref{Oi_def}) already have very good overlap with $\Upsilon(nS)$ states. 
Therefore, we choose $\tau_0=0$ in this study. It has been
checked in our previous work that using larger values of $\tau_0$ does not
change the results significantly \cite{Larsen:2019zqv}.
To obtain the BS amplitude, we consider the large $\tau$ behavior of
the following combination:
\begin{equation}
e^{E_{\alpha} \tau} \tilde C^r_{\alpha}(\tau)=e^{E_{\alpha} \tau} \sum_j \Omega_{j \alpha} \expval{O_{qq}^r(\tau) O_j(0)} .
\end{equation}
\begin{figure}[t!]
\includegraphics[width=8cm]{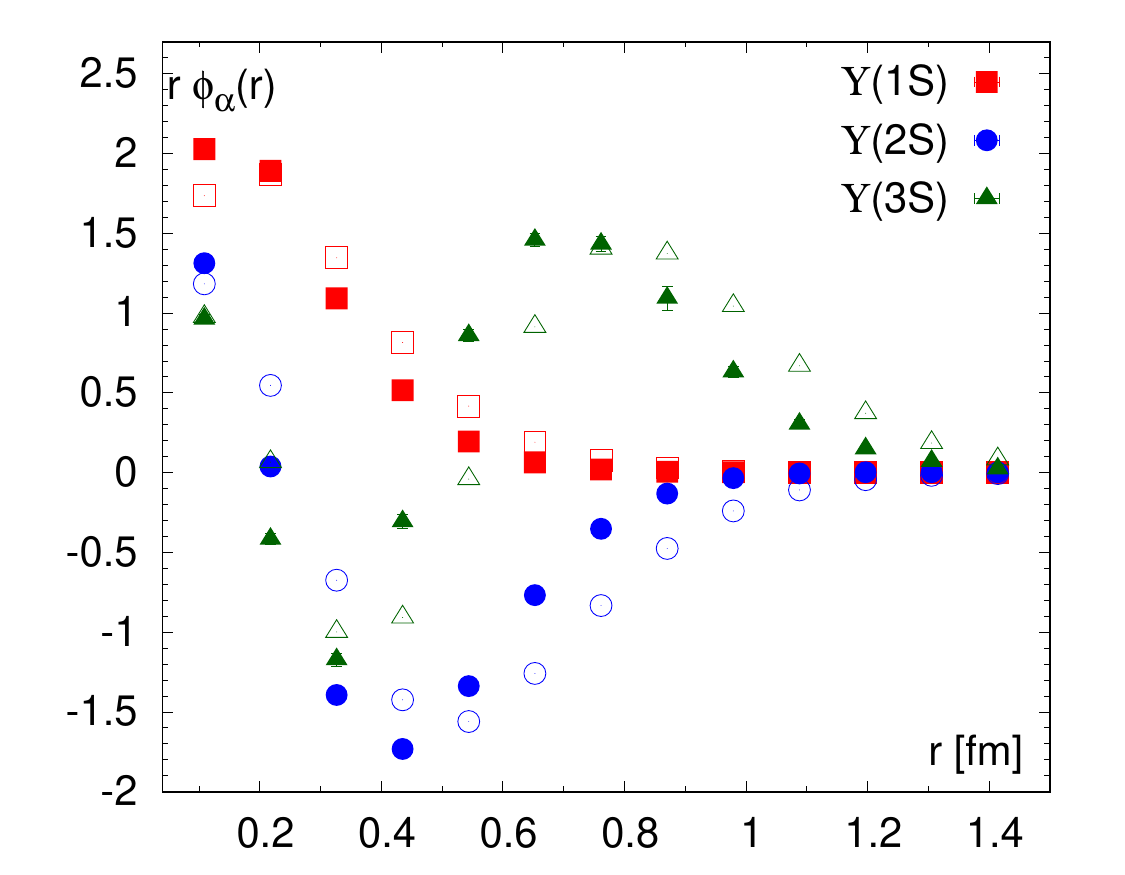}
\caption{The BS amplitudes for $\Upsilon(1S)$, $\Upsilon(2S)$, and $\Upsilon(3S)$ states at \(T=0\)
as function of $r$ (filled symbols) compared with the corresponding trial wave functions (open symbols).}
\label{fig:BS_Ups}
\end{figure}
The energy $E_{\alpha}$ has been determined from the fits of the correlators of the optimized operators $\tilde O_{\alpha}$ \cite{Larsen:2019zqv}.
In practice, the value of $\tau$ does not have to be very large.  We find that
$\tau>0.3$ fm works for all states; i.e., the resulting BS amplitudes are time
independent. For $\tau=0$ the BS amplitude will be equal to the trial wave function
$\psi_i(r)$. To obtain the proper normalization of the BS amplitude we require that
$\int_0^{\infty} dr r^2 |\phi_{\alpha}(r)|^2=1$. After this normalization exponential
factor $e^{E_{\alpha} \tau}$ drops out. Therefore, the normalized BS amplitudes do
not depend on the choice of the energy $E_{\alpha}$.
In Fig. \ref{fig:BS_Ups} we show the
BS amplitude $\phi_{\alpha}(r)$ for $\Upsilon(1S)$, $\Upsilon(2S)$ and $\Upsilon(3S)$
states compared to the corresponding trial wave functions $\psi_{\alpha}(r)$ used to
construct the optimized meson operators. We see that the $r$-dependence of the BS
amplitudes is in qualitative agreement with the expectations of non-relativistic
potential model. However, the details of the $r$ dependence are different from the
input trial wave function. We also note that the orthogonalization procedure is
important for getting the correct $r$ dependence of the BS amplitudes.

\begin{figure}[t!]
\includegraphics[width=8cm]{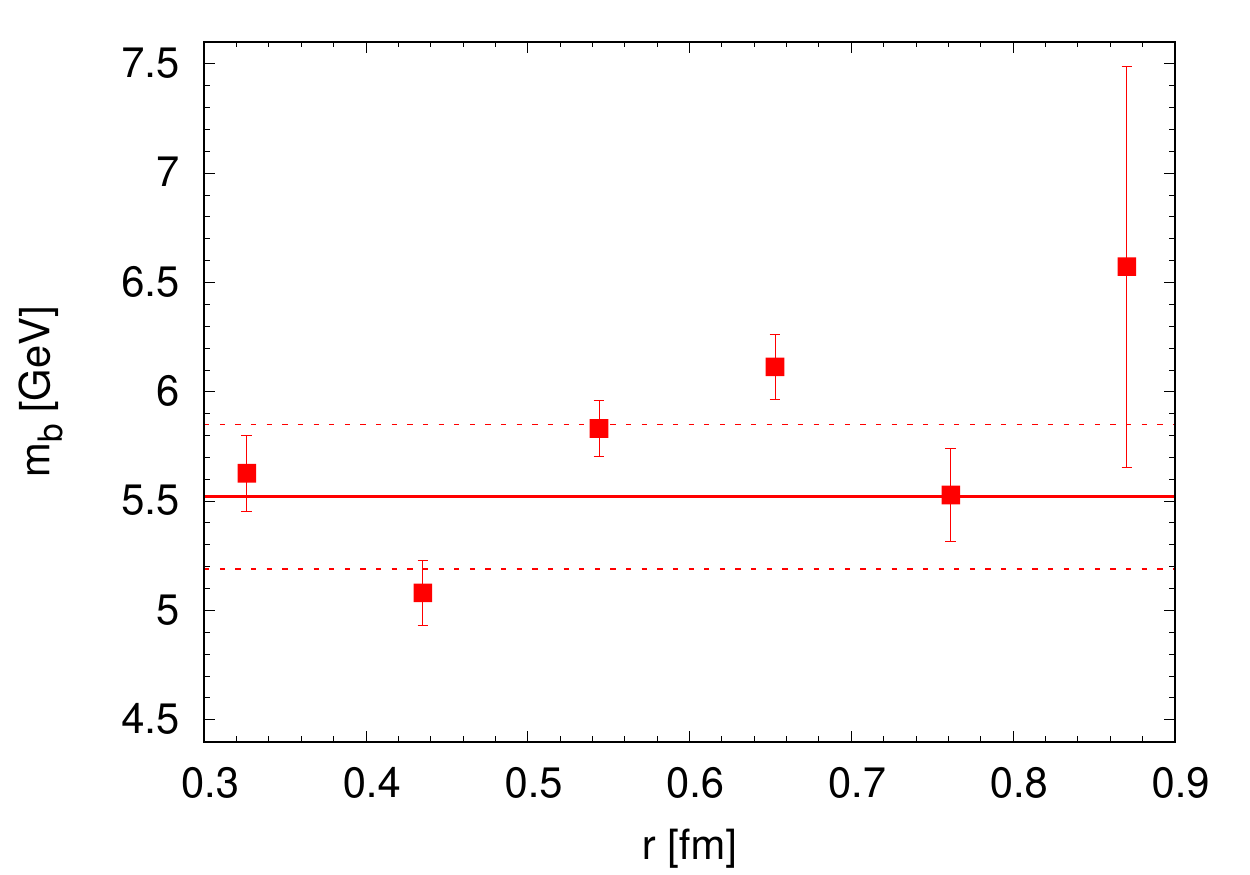}
\caption{The effective bottom quark mass, $m_b$, in the potential approach determined
for different quark antiquark separations $r$ (see text). The horizontal solid line
is the fitted value of $m_b$, while the dashed lines indicate the corresponding
uncertainty.}
\label{fig:mb}
\end{figure}

If the potential picture is valid the BS amplitude should satisfy the Schr\"odinger equation
\begin{equation}
\left( \frac{-\nabla^2}{m_b} + V\left(r\right) \right)\phi_{\alpha}=E_{\alpha} \phi_{\alpha},
\end{equation}
with $m_b$ being the b-quark mass of the potential model. Note that the reduced mass in the $b \bar b$ system is $m_b/2$, hence
the absence of factor two in the above equation.
Using the BS amplitude and the energy of  at least two bottomonia states determined in NRQCD from the above equation we can obtain $m_b$
and the potential $V(r)$. We determine the $b$-quark mass using $\Upsilon(1S)$ and $\Upsilon(2S)$ states as follows
\begin{equation}
m_b=\frac{\displaystyle \frac{\nabla^2 \phi_{\Upsilon(1S)}}{\phi_{\Upsilon(1S)}}-\frac{\nabla^2 \phi_{\Upsilon(2S)}}{\phi_{\Upsilon(2S)}}}{E_{\Upsilon(2S)}-
E_{\Upsilon(1S)}}
\end{equation}
To evaluate $\nabla \phi_{\alpha}$ we use the simplest difference scheme. The value of $m_b$ determined from the above equation
for different values of
quark antiquark separation $r$ is shown in Fig. \ref{fig:mb}.
The $r$-range was chosen such that it does not include the node of $\Upsilon(2S)$ and large distances, where the statistical errors
are large.
We see some modulation of the extracted $m_b$ in $r$, which may indicate that the BS
amplitude cannot be completely captured by the Schr\"odiner equation, but there is no clear tendency of $m_b$ as function of $r$.
Therefore we fitted the values of $m_b$ obtained for different $r$ to a constant. This resulted in
\begin{equation}
m_b=5.52 \pm 0.33 ~{\rm GeV}.
\end{equation}
This value of the effective bottom quark mass obtained by us is not very different from the one used by the original Cornell model,
$m_b=5.17$ GeV~\cite{Eichten:1979ms}, but is significantly larger than the $b$-quark mass used in most of the potential models (see, e.g.,
Ref. \cite{Jacobs:1986gv}). We also determined the value of $m_b$ using the BS amplitudes and the energy levels of
$\Upsilon(1S)$ and $\Upsilon(3S)$ and obtained $m_b=5.82(0.51)$ GeV. This agrees with the above result within the errors.

\begin{figure}[t!]
\includegraphics[width=8cm]{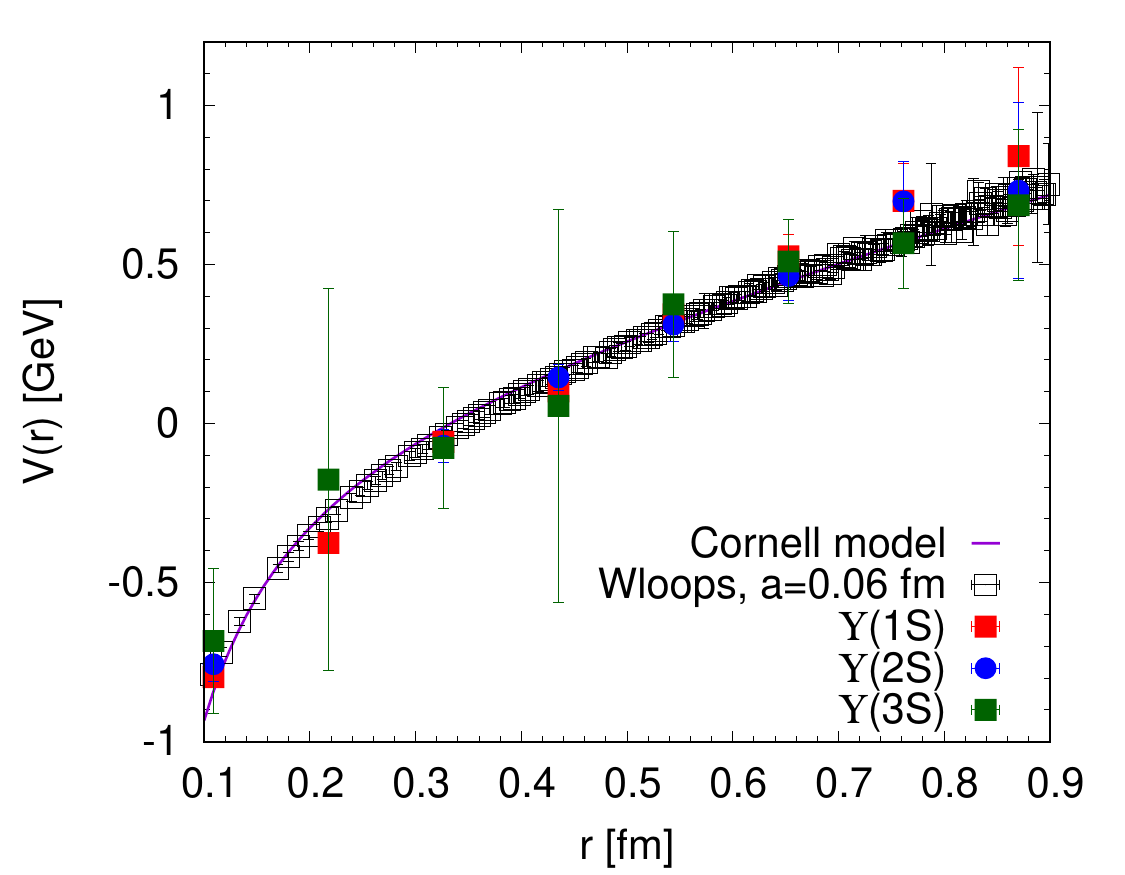}
\caption{The potential, $V(r)$, obtained from the BS amplitude of $\Upsilon(1S)$,
$\Upsilon(2S)$ and $\Upsilon(3S)$ states compared to the phenomenological Cornell
potential~\cite{Eichten:1979ms} shown as a solid line
as well as to the the energy of the static quark antiquark pair obtained from Wilson
loops using $a=0.06$~fm lattice~\cite{Bazavov:2014pvz}. All the lattice results were
normalized to coincide with the Cornell potential at $r=0.4$~fm.}
\label{fig:pot}
\end{figure}

Having determined $m_b$, we can also calculate the potential, $V(r)$, using the BS amplitudes and the bottomonium energy
levels as
\begin{equation}
V(r)=E_{\alpha} + \frac{\nabla^2 \phi_{\alpha}}{m_b \phi_{\alpha}}.
\end{equation}
The results are shown in Fig.~\ref{fig:pot}. Given our findings for $m_b$, it is not
surprising that the values of the potential obtained using $\Upsilon(1S)$,
$\Upsilon(2S)$, and $\Upsilon(3S)$  states agree within errors. In the figure, we also
compare the value of $V(r)$ determined from the different states to the
phenomenological potential of the original Cornell model~\cite{Eichten:1979ms}
and the energy of static quark antiquark pair
obtained from Wilson loops at lattice spacing $a=0.06$ fm~\cite{Bazavov:2014pvz}. It
is quite non-trivial that all these potentials agree with each other. A similar
conclusion is reached in
Refs.~\cite{Kawanai:2011xb,Kawanai:2011jt,Kawanai:2013aca} when the limit of quark
mass going to infinity is taken. We note that the relativistic corrections to the
spin-dependent part of the potential are quite small for the $b$ quark mass \cite{Bali:1997am} 
and thus are not visible given our statistical errors.

The discussion above ignored spin-dependent effects. To address the spin-dependent
part of the potential we also calculated the BS amplitude for $\eta_b(nS)$ states,
$n=1,2,3$. We have found that the corresponding BS amplitudes agree with the ones of
the $\Upsilon(nS)$ states within errors. Therefore, with the present statistics, we
cannot resolve the spin-dependent part of the potential.

\begin{figure}[t!]
\includegraphics[width=8cm]{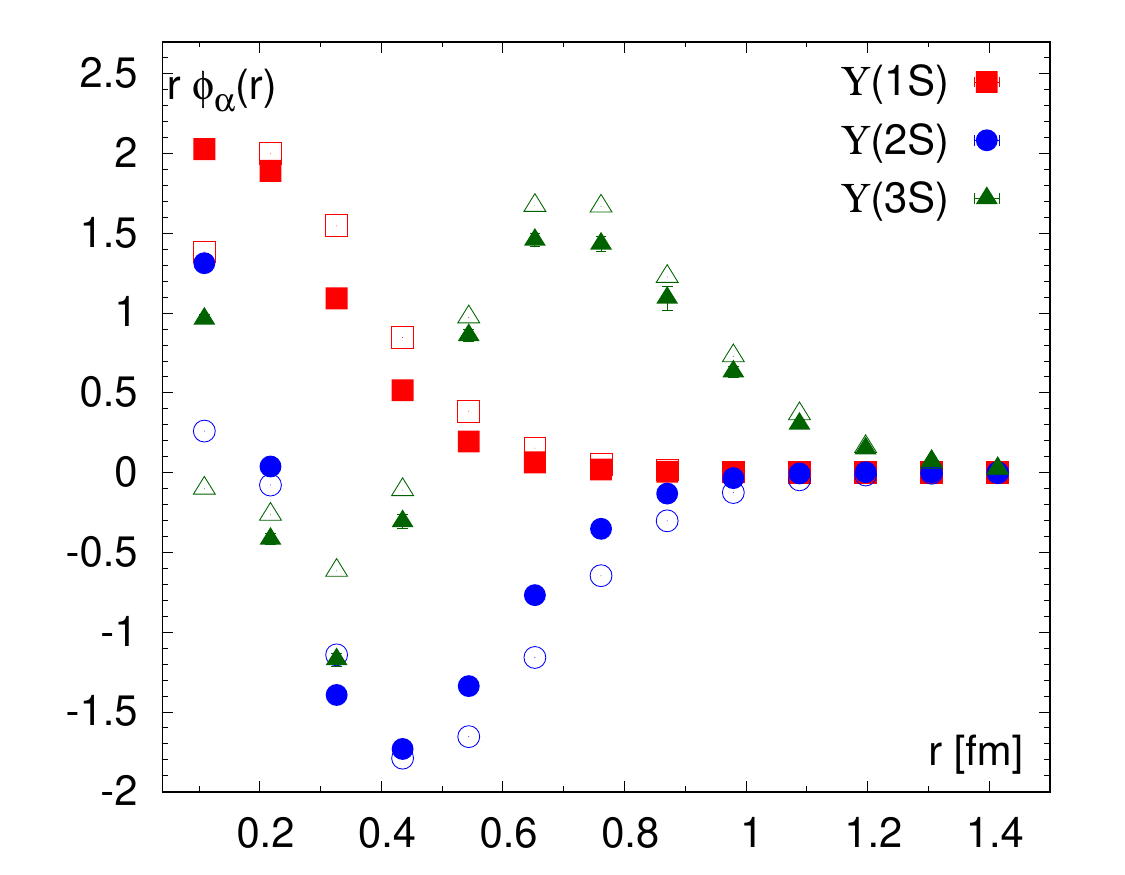}
\caption{The BS amplitude for $\Upsilon(nS)$ states as function of $r$ (filled symbols) compared
with the non-relativistic wave functions obtained from potential model with $m_b=6$ GeV (open symbols).}
\label{fig:wf}
\end{figure}

As discussed above, the $r$-dependence of the BS amplitudes qualitatively follow the
$r$-dependence of the trial wave function $\psi_i(r)$ obtained from the potential
model. But at qualitative level, significant differences can be seen, (cf.
Fig.~\ref{fig:BS_Ups}). This potential model used $m_b=4.676$
GeV~\cite{Meinel:2010pv}, which is smaller than the effective quark masses determined
above. Therefore, we calculated the wave functions of $(nS)$ bottomonium states using
the static quark anti-quark energy~\cite{Bazavov:2014pvz} as a potential and $m_b=6$ GeV. The results are shown in
Fig.~\ref{fig:wf} and we see that the agreement between the BS amplitude and the wave
functions is significantly improved. We also note that the dependence of the energy
levels on $m_b$ is rather mild; e.g., changing $m_b$ from $4$ to $6$ GeV only
reduces the spin-averaged 2S-1S splitting  by $3.5\%$. Thus, using large values of
$m_b$ in the potential model is a viable option.

\section{Bethe-Salpeter amplitudes at \(\mathbf{T>0}\)}

\begin{figure*}
\includegraphics[width=8cm]{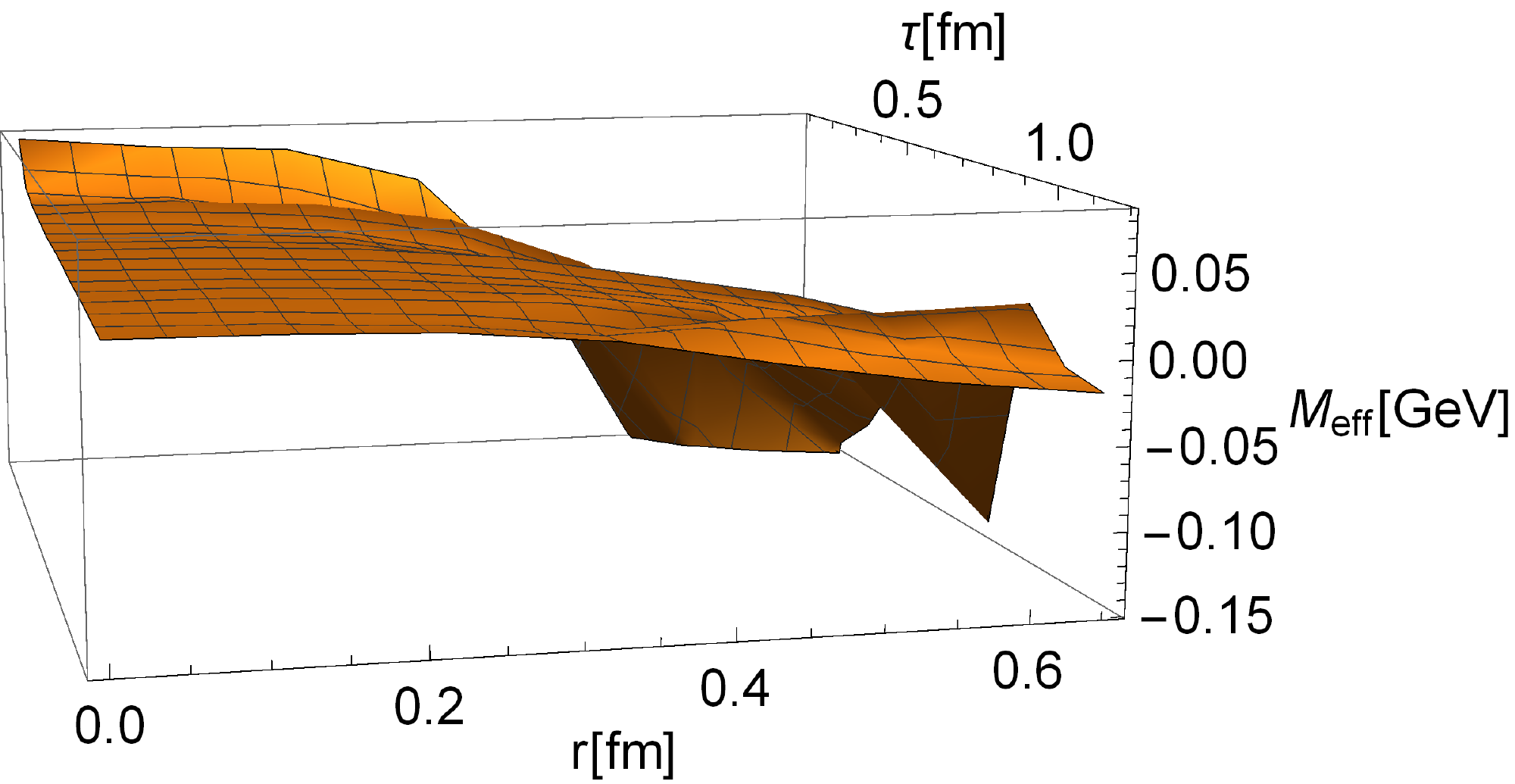}
\includegraphics[width=8cm]{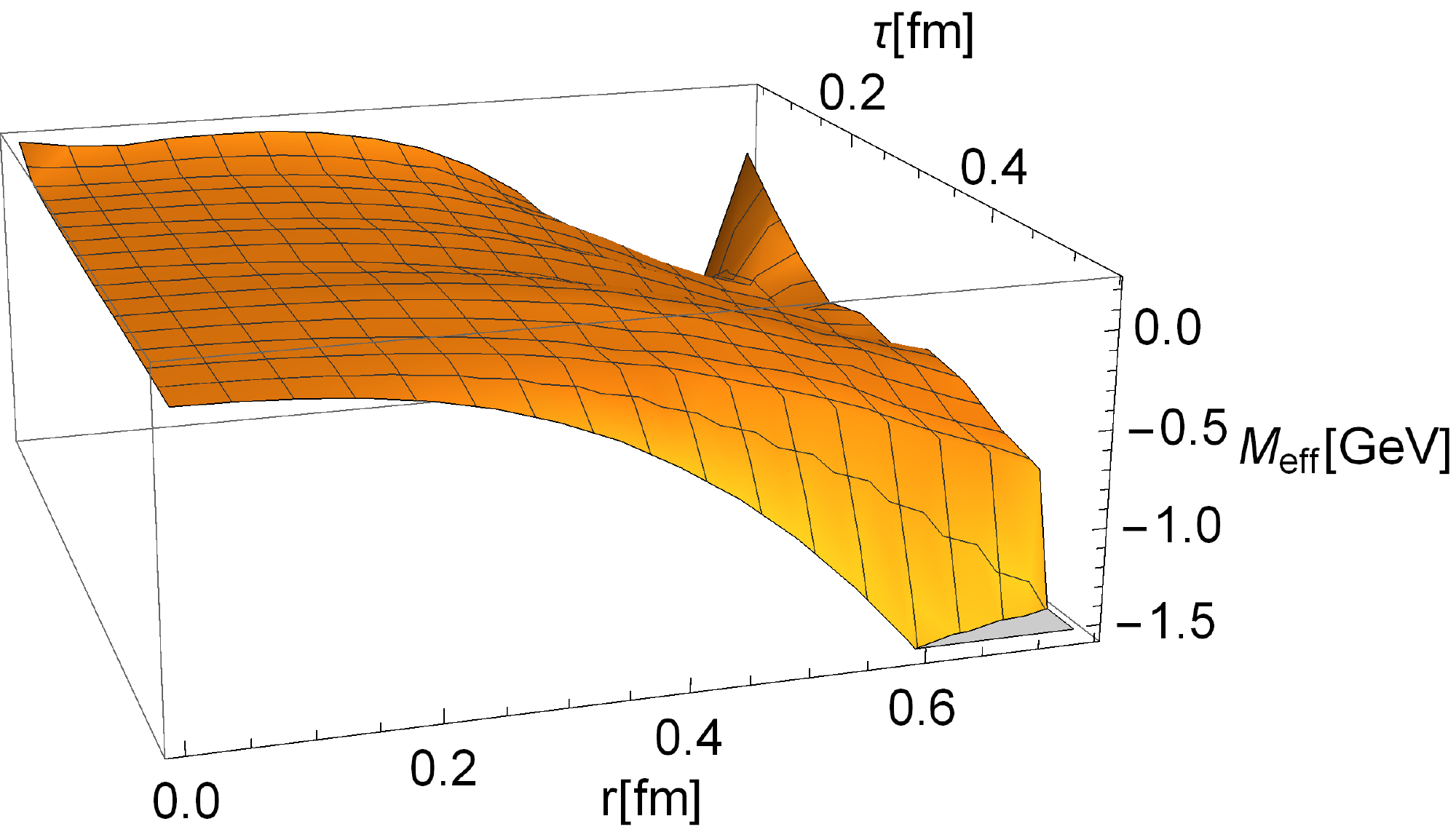}
\caption{The effective masses $M_{\mathrm{eff}}^r(\tau,T)$ in GeV of the $\Upsilon(1S)$ correlator at $T=151$ MeV (left)
and $T=334$ MeV (right) as function of $\tau$ and $r$.}
\label{fig:effmass_1S}
\end{figure*}

We can also consider the mixed correlator $\tilde C^r_{\alpha}(\tau,T)$ defined in
Eq.~(\ref{def_Calpha}) for \(T>0\) by evaluating the expectation value over a
thermal ensemble at a temperature \(T=1/\beta\),
\begin{equation}
  \tilde C^r_{\alpha}(\tau,T) = \frac{1}{Z(\beta)} \Tr \qty[ O^r_{qq}(\tau) \tilde
  O_{\alpha}(0) e^{-\beta H} ] ,
  \label{def_Calpha_T}
\end{equation}
with the thermal partition function \(Z(\beta)=\Tr \qty[e^{-\beta H}]\). Using energy
eigenstates to evaluate the trace and inserting a complete set of states we obtain
the following expression for the correlator $\tilde C^r_{\alpha}(\tau,T)$:
\begin{equation}
  \begin{split}
    & \tilde C^r_{\alpha}(\tau,T) = \\
    & \frac{1}{Z(\beta)} \sum_{n,m} e^{-(E_n-E_m)\tau }
    \mel{m}{O_{qq}^r}{n} \mel{n}{\tilde O_{\alpha}}{m} e^{-\beta E_m} .
  \end{split}
\end{equation}
Since we perform calculations in NRQCD, the sum over $m$ should be restricted to states that do no contain
the heavy quark anti-quark pair; heavy quark pair creation is not allowed in NRQCD.
We denote those states as $\ket{m^\prime}$. If we write the states $\ket{n}$ as $\ket{n^\prime \gamma}$, where
index $n'$ labels the light degrees of freedom and $\gamma$ labels the quarkonium states,
the above expression for $\tilde C^r_{\alpha}(\tau,T)$ can be rewritten as
\begin{equation}
  \begin{split}
  \tilde C^r_{\alpha}(\tau,T) = \frac{1}{Z(\beta)} &  \sum_{\gamma,n^\prime,m^\prime}
    \left[ e^{-(E_{n^\prime,\gamma}-E_m^\prime)\tau} e^{-\beta E_{m^\prime}} \right. \\
    & \left. \mel{m^\prime}{O_{qq}^r}{n^\prime \gamma} \mel{n^\prime \gamma}{\tilde O_{\alpha}}{m^\prime}
    \right ] .
  \end{split}
\end{equation}
If we write $E_{m^\prime \gamma}=E_{\gamma}+E_{m^\prime}+\Delta E_{m^\prime \gamma}$
and assume that the operator $\tilde O_{\alpha}$ mostly projects onto quarkonium
state $\ket{\alpha}$, we can obtain a simplified form,
\begin{equation}
  \begin{split}
  & \tilde C^r_{\alpha}(\tau,T) = e^{-E_{\alpha} \tau} \left[ \phi_{\alpha} A_{\alpha}^{*^\prime}
   + \frac{1}{Z(\beta)} \times \right. \\
   &  \left.  \sum_{m^\prime}
   \mel{m^\prime}{O_{qq}^r}{m^\prime \alpha} \mel{m^\prime \alpha}{\tilde O_{\alpha}}{m^\prime}
  e^{ -\beta E_{m^\prime} - \Delta E_{m^\prime \alpha} \tau}  \right ]
  \end{split}
  \label{decomp_final}
\end{equation}
with $A_{\alpha}^{*^\prime}=A_{\alpha}^{*}/Z(\beta)$. In the above equation, we
separated out the the $m^\prime=0$ vacuum contribution in the sum corresponding to
the thermal trace. At small temperature, the first term in the above equation is the
dominant one, and the correlator is approximately given by the $T=0$ BS amplitude. In
general, however, there is no simple interpretation of the correlator $\tilde
C_{\alpha}^r(\tau,T)$ in terms of some finite temperature quarkonium wave function.
The temperature dependence of this correlator crucially depends on the value of the
matrix elements $\mel{m^\prime}{O_{qq}^r}{m^\prime \alpha}$ and $\mel{m^\prime
\alpha}{\tilde O_{\alpha}}{m^\prime}$. The size of $\mel{m^\prime}{O_{qq}^r}{m^\prime
\alpha}$ depends on the separation $r$ and therefore, also the size of the thermal
effect will be $r$ dependent. For values of $r$ that are about the size of the
bottomonium state of interest the matrix elements  $\mel{m^\prime}{O_{qq}^r}{m^\prime
\alpha}$ and $\mel{m^\prime \alpha}{\tilde O_{\alpha}}{m^\prime}$ should be of
similar size, and thus the temperature dependence of $\tilde C_{\alpha}^r(\tau,T)$ is
expected to be comparable to the correlator of $\tilde O_{\alpha}$ explored in
Ref.~\cite{Larsen:2019zqv}.

We performed calculations of $\tilde C_{\alpha}^r$ at six different temperatures using $48^3 \times 12$
lattices from HotQCD collaboration. The parameters of the calculations including the gauge
coupling $\beta=10/g_0^2$ and number of configurations are summarized in Table \ref{tab:par}.
As at zero temperature, we used 8 sources per gauge configuration. The projection matrix
$\Omega_{j \alpha}$ has been determined from the finite temperature correlators according
to Eq. (\ref{Omega_def}). We checked, however, that the difference between the finite temperature
projection matrix and the zero temperature projection matrix is very small.
\begin{table}
\begin{tabular}{ccc}
\hline
$\beta$ & $T$ (MeV) & Number of configs. \\
\hline
6.740 & 151 & 384  \\
6.880 & 172 & 384  \\
7.030 & 199 & 384  \\
7.280 & 251 & 384  \\
7.596 & 334 & 384  \\
\hline
\end{tabular}
\caption{The parameters for the \(2+1\) flavor HISQ ensembles at \(T>0\) with
\(48^3\times12\) lattices.}
\label{tab:par}
\end{table}

\begin{figure*}[t!]
  \includegraphics[width=8cm]{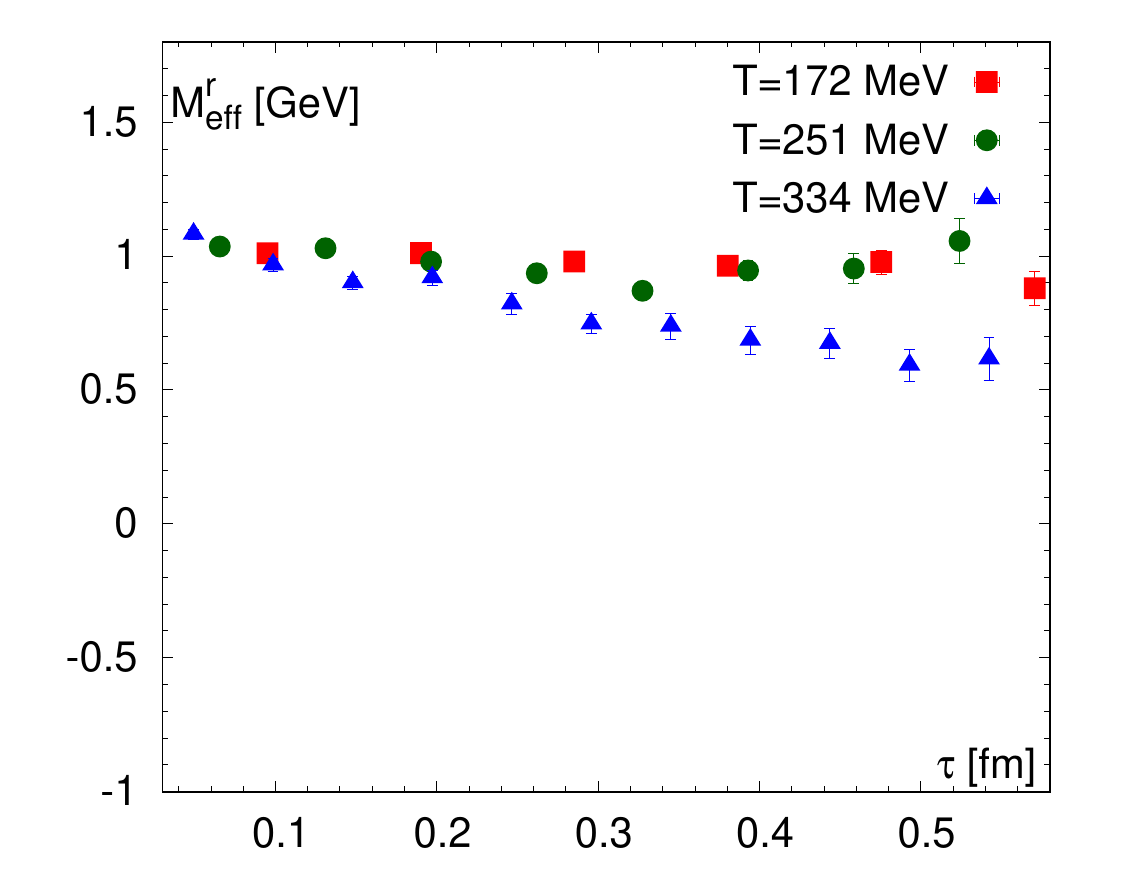}
  \includegraphics[width=8cm]{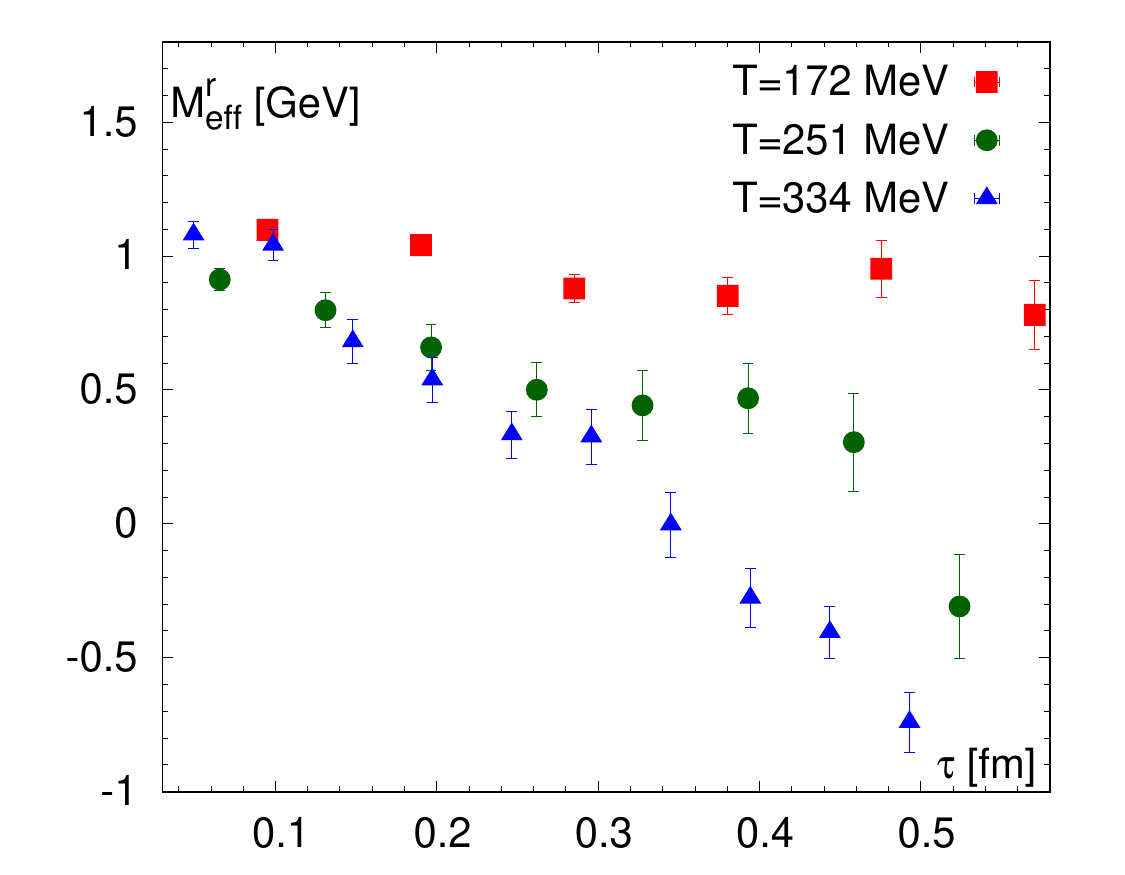}
\caption{The effective masses $M_{\mathrm{eff}}^r(\tau,T)$ in GeV of the $\Upsilon(3S)$ correlator for $r \simeq 0.25$ fm (left)
and $r \simeq 0.65$ fm (right) at different temperatures as function of $\tau$.}
\label{fig:effmass_3S}
\end{figure*}

We could use the same approach as in Ref. \cite{Larsen:2019zqv} to explore the temperature
dependence of the correlator $\tilde C_{\alpha}^r(\tau,T)$  and define
the effective mass for a fixed \(r\),
\begin{equation}
  a M_{\mathrm{eff}}^r(\tau,T)=\ln( \frac{\tilde C_{\alpha}^r\left(\tau,T\right) }{\tilde
  C_{\alpha}^r\left(\tau+a,T)\right)} ).
\end{equation}
Now, the effective mass also depends on the distance $r$ between the quark and
antiquark in the point-split current. In Fig. \ref{fig:effmass_1S}, we show the
effective mass of $\Upsilon(1S)$ correlator as function of $r$ and $\tau$ at the
lowest and the highest temperature. The errors of the effective masses are not shown
to improve the visibility. Since the energy levels in NRQCD are only defined up to a
lattice spacing dependent constant, as in Ref.~\cite{Larsen:2019bwy}, we calibrate the
effective masses with respect to the energy level of $\eta_b(1S)$ state at zero
temperature. At large $\tau$ and $r$, the errors are quite large, and within these
errors we do no see any medium effects in the effective mass at the lowest
temperature. For small $r$, the effective mass quickly reaches  a plateau with
increasing $\tau$. For large $r$, the effective mass at $151$ MeV reaches the plateau
from below. At the highest temperature, $T=334$ MeV, the $r$ and $\tau$ dependences of
the effective masses looks similar for not too large values of $r$. However, the
behavior of the effective mass is qualitatively different for large $r$. In
particular, the effective mass does not reach a plateau with increasing $\tau$. For
excited states, the results for $M_{\mathrm{eff}}^r(\tau,T)$ look similar, except that the
errors are very large for $r>0.65$ fm. As an example, we show the effective mass for
$\Upsilon(3S)$ in Fig.~\ref{fig:effmass_3S} at two values of $r$, $r=0.25$ fm and
$r=0.65$ fm for different temperatures. For the smaller $r$, we see no temperature
dependence of the $\Upsilon(3S)$  effective mass at $T=172$ MeV and $T=251$ MeV. This
is likely due to the fact that the matrix elements $\mel{m^\prime}{O_{qq}^r}{
m^\prime\Upsilon(3S)}$ are small for $r=0.25$~fm and the first term in Eq.~(\ref{decomp_final}) 
dominates. Note, however, that the errors are large. For the
highest temperature, $T=334$ MeV we start to see significant temperature dependence.
For the larger distance, $r=0.65$ fm, the medium effects are more pronounced. While
the modifications of $M_{\mathrm{eff}}^r$ are small for $T=172$ MeV, thermal effects are
significant for $T=251$ MeV and $334$ MeV, comparable in size to the thermal effects
in the effective masses of correlators of optimized operators \cite{Larsen:2019zqv}.

\begin{figure}[t!]
\includegraphics[width=8cm]{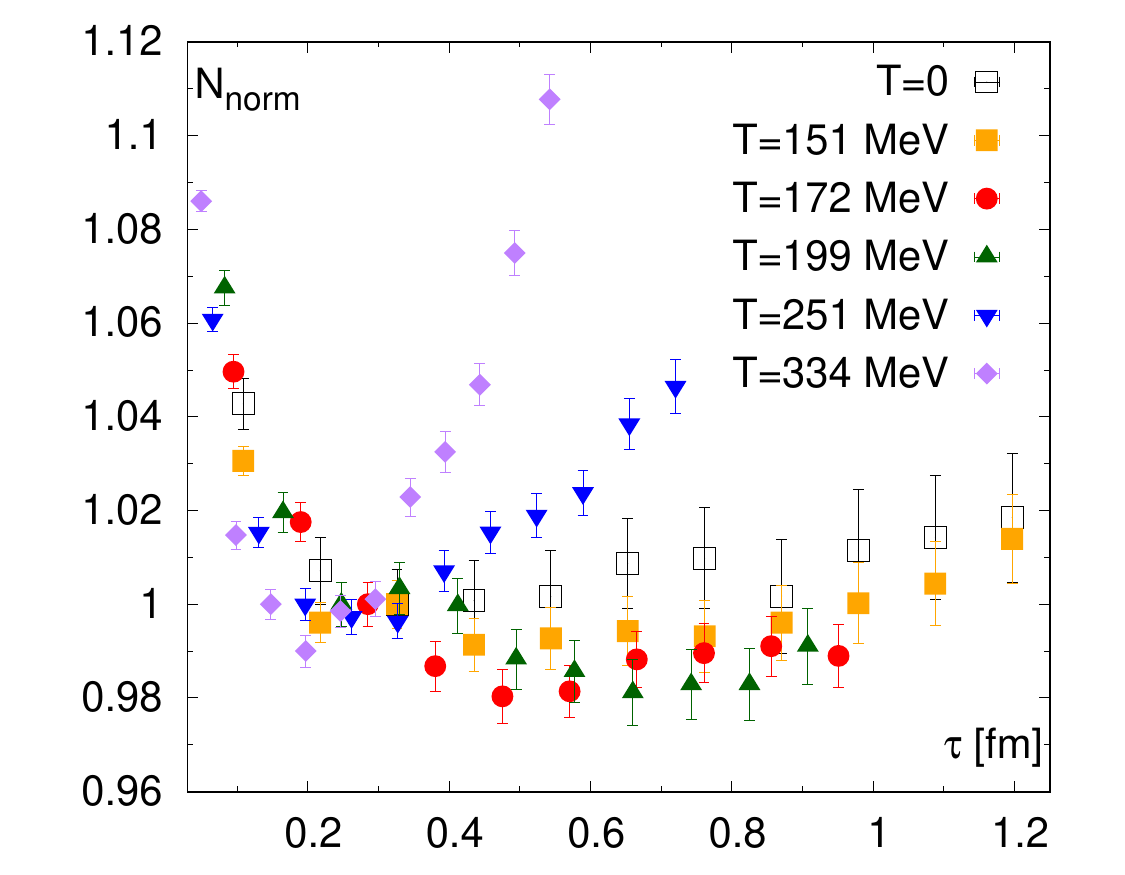}
\includegraphics[width=8cm]{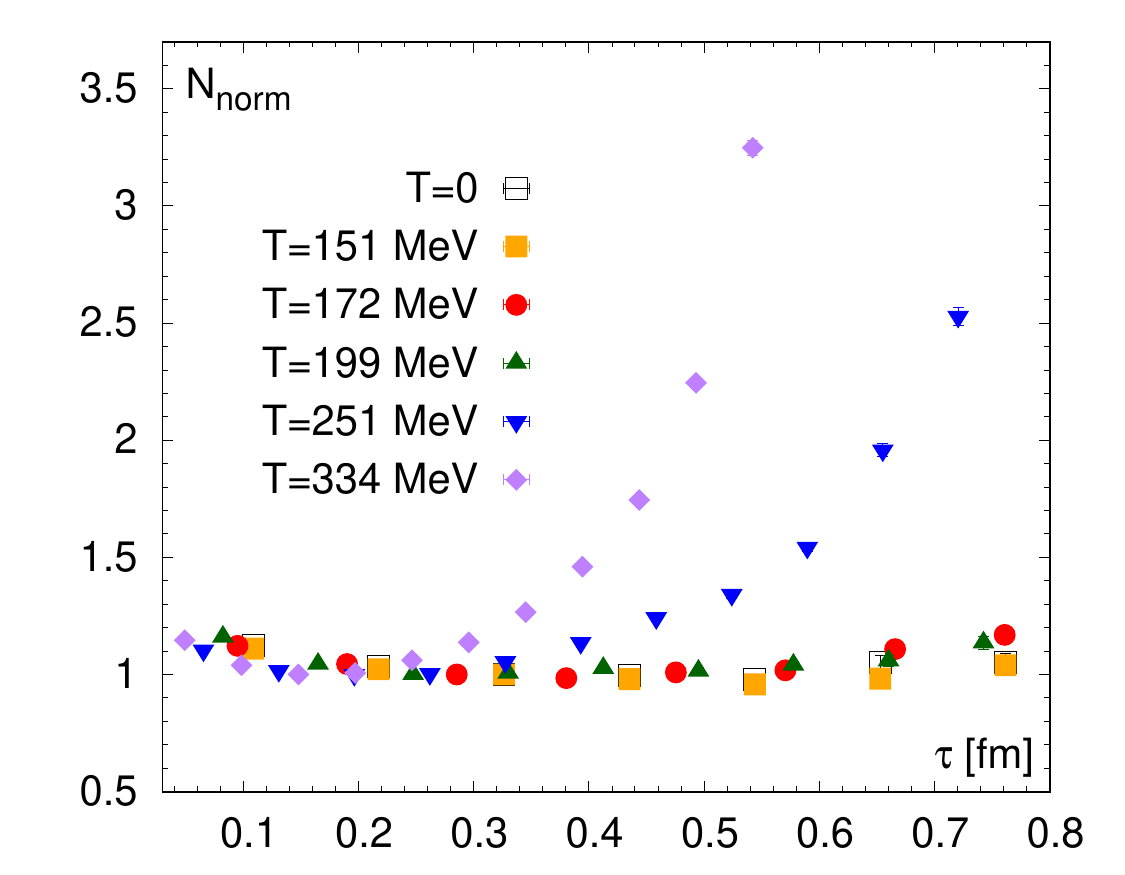}
\includegraphics[width=8cm]{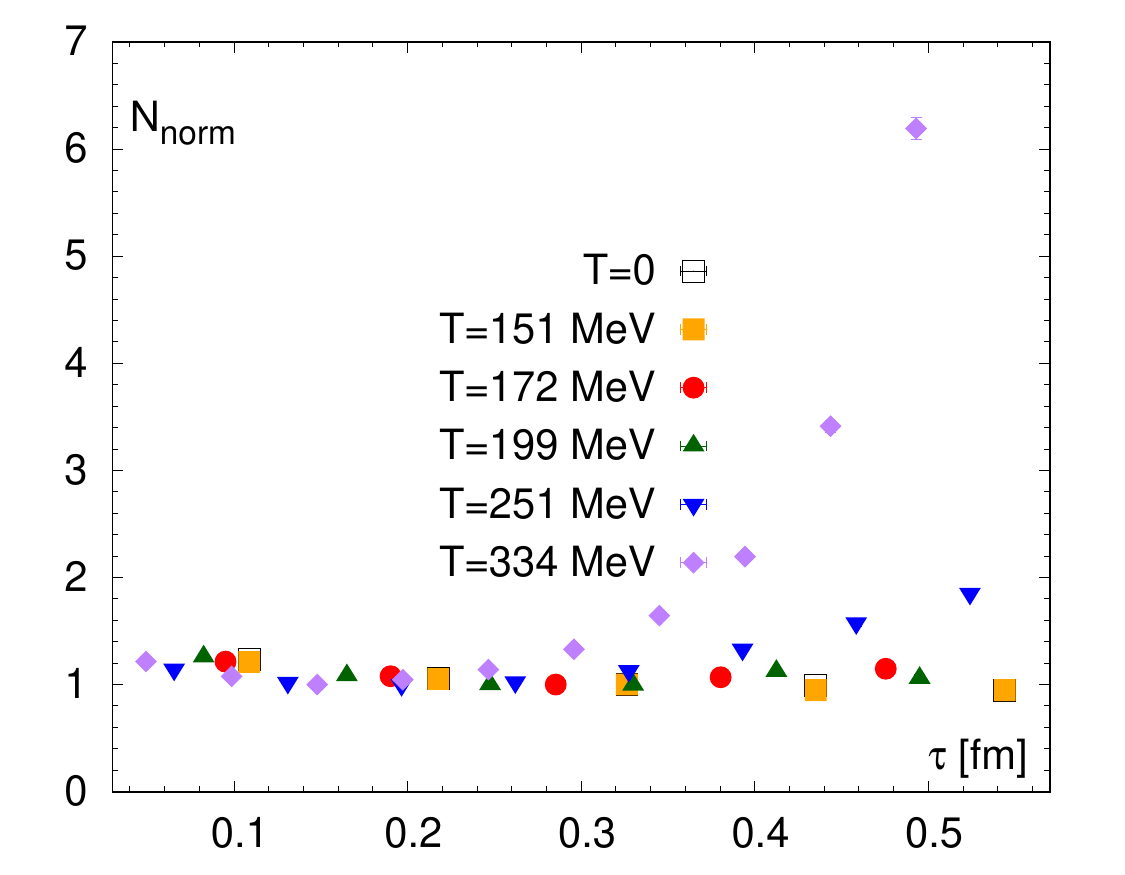}
\caption{Norm  of the squared BS wave function at different temperatures for the $\Upsilon(1S)$ (Top),
$\Upsilon(2S)$ (Middle) and $\Upsilon(3S)$ (Bottom) states. }
\label{fig:Norm}
\end{figure}

Since the correlator $\tilde C_{\alpha}^r$ does not correspond to a positive definite spectral
function, it is difficult to infer in-medium properties of bottomonia from $M_{\mathrm{eff}}^r$.
The large statistical errors make this even more complicated. Another way to analyze the
temperature dependence of $\tilde C_{\alpha}^r$ is to consider the integral
\begin{equation}
N_{\alpha}(\tau,T)=\int_0^{\infty} dr r^2 \Big(\tilde C_{\alpha}^r\Big)^2.
\end{equation}
At zero temperature, this quantity should be proportional to $\exp(-2 E_{\alpha} \tau)$ for
sufficiently large $\tau$. This is also expected to be true below the crossover temperature.
The combination
\begin{equation}
N_{\mathrm{norm}}(\tau,T)=\exp(2 E_{\alpha} \tau) N_{\alpha}(\tau,T)
\end{equation}
should be independent of $\tau$ and
can be interpreted as the normalization of the BS amplitude. In Fig. \ref{fig:Norm}, we
show $N_{\mathrm{norm}}(\tau,T)$ as function of $\tau$ for different temperatures normalized to one
at $t=\tau/a=3$. As before, the energy values, $E_{\alpha}$, have been determined from
the correlators of optimized operators at $T=0$ \cite{Larsen:2019zqv}.

For the lowest temperature as well as for $T=0$, we see that $N_{\mathrm{norm}}(\tau,T)$
is approximately constant as expected. Here, we note that the $\tau$ range in Fig. \ref{fig:Norm}
is different for $\Upsilon(1S)$, $\Upsilon(2S)$, and $\Upsilon(3S)$ states. This is due to
the fact that the correlators $C_{\Upsilon(2S)}^r$ and $C_{\Upsilon(3S)}^r$ will be contaminated
by the lowest $\Upsilon(1S)$ state at large $\tau$ as the projection is not perfect due to
the small operator basis of only three operators used in this study.
As the temperature increases we see that $N_{\mathrm{norm}}(\tau,T)$ no longer
approaches a constant but increases at large $\tau$.
This implies that the correlator $\tilde C_{\alpha}^r$ is no longer dominated by the first term
in Eq. (\ref{decomp_final}).
The $\tau$-dependence of $N_{\mathrm{norm}}(\tau,T)$
is larger for high temperatures and is also more pronounced for excited states, as expected.
\begin{figure}[h]
\includegraphics[width=8cm]{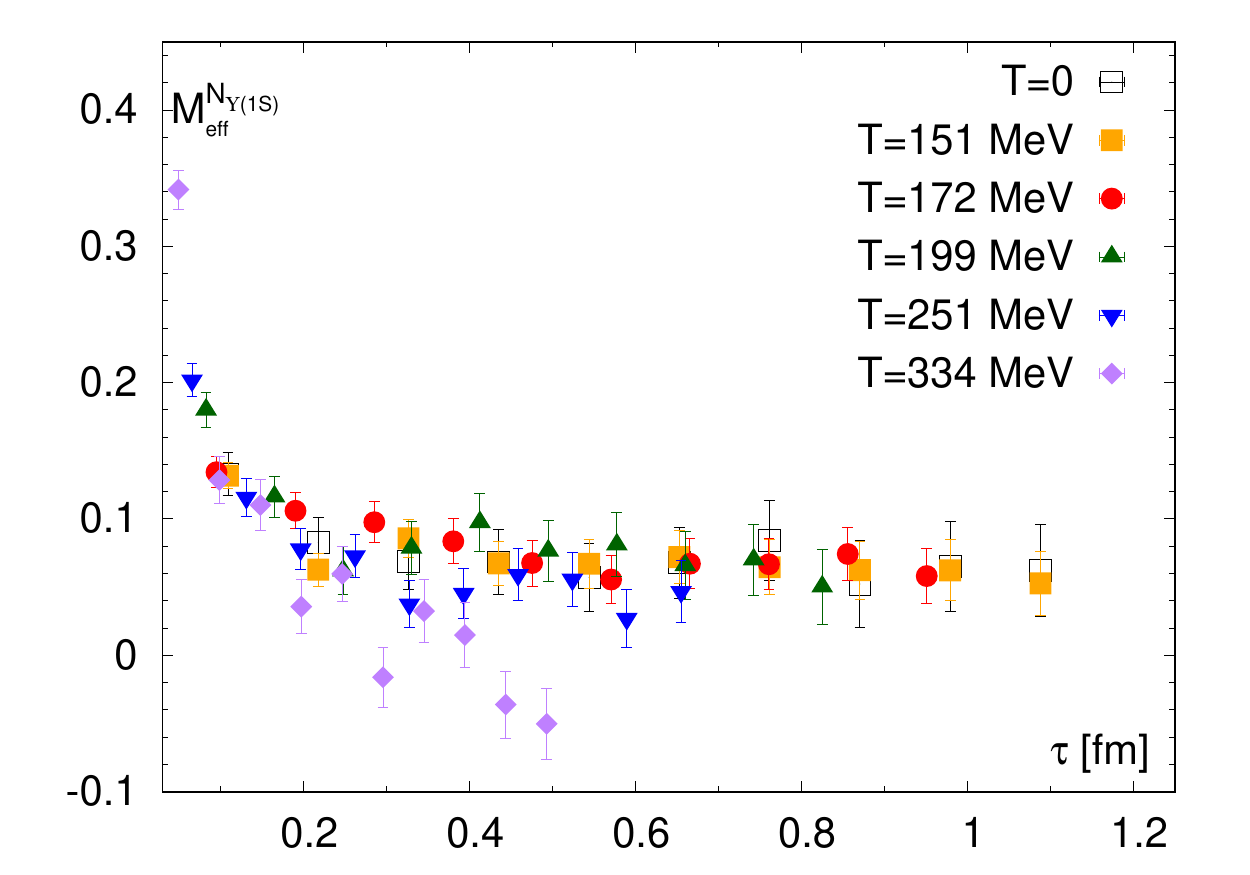}
\includegraphics[width=8cm]{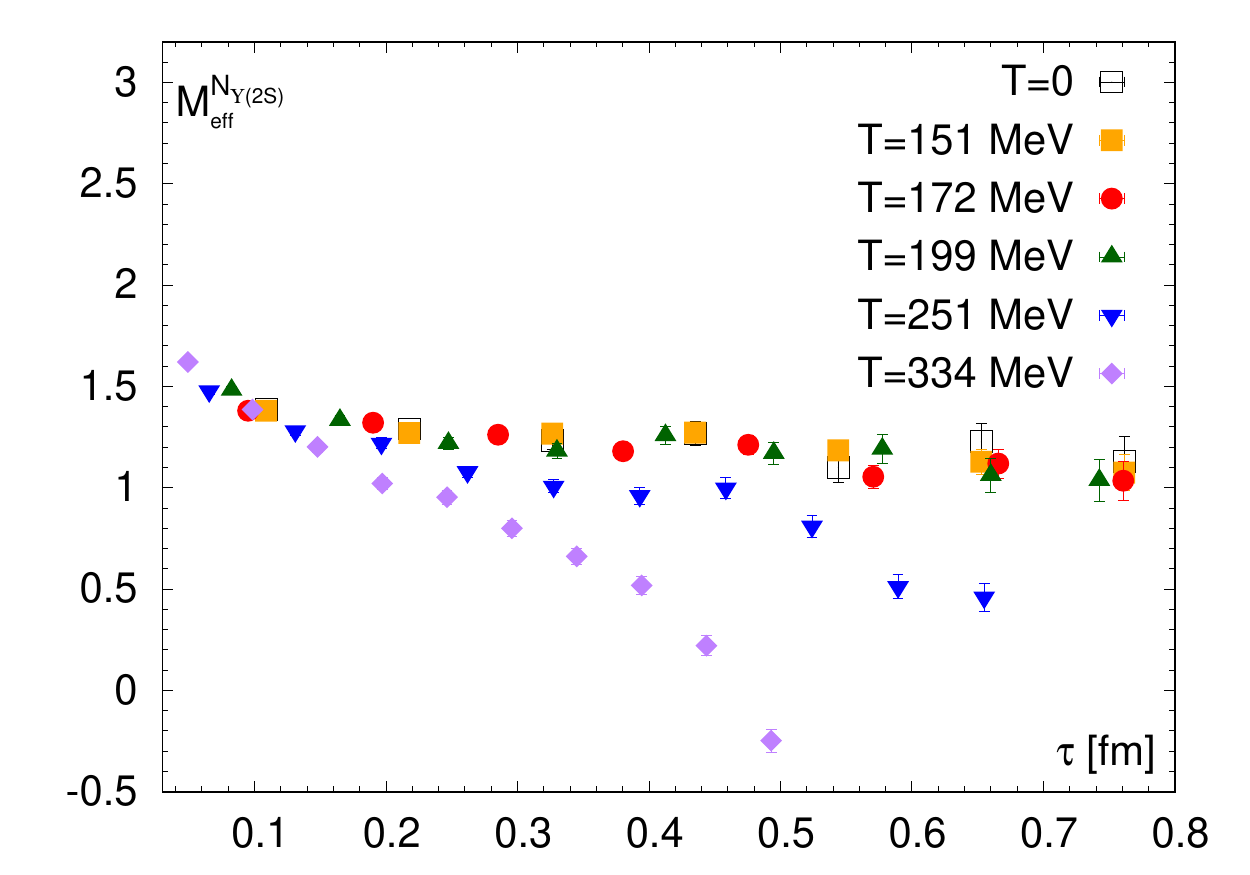}
\includegraphics[width=8cm]{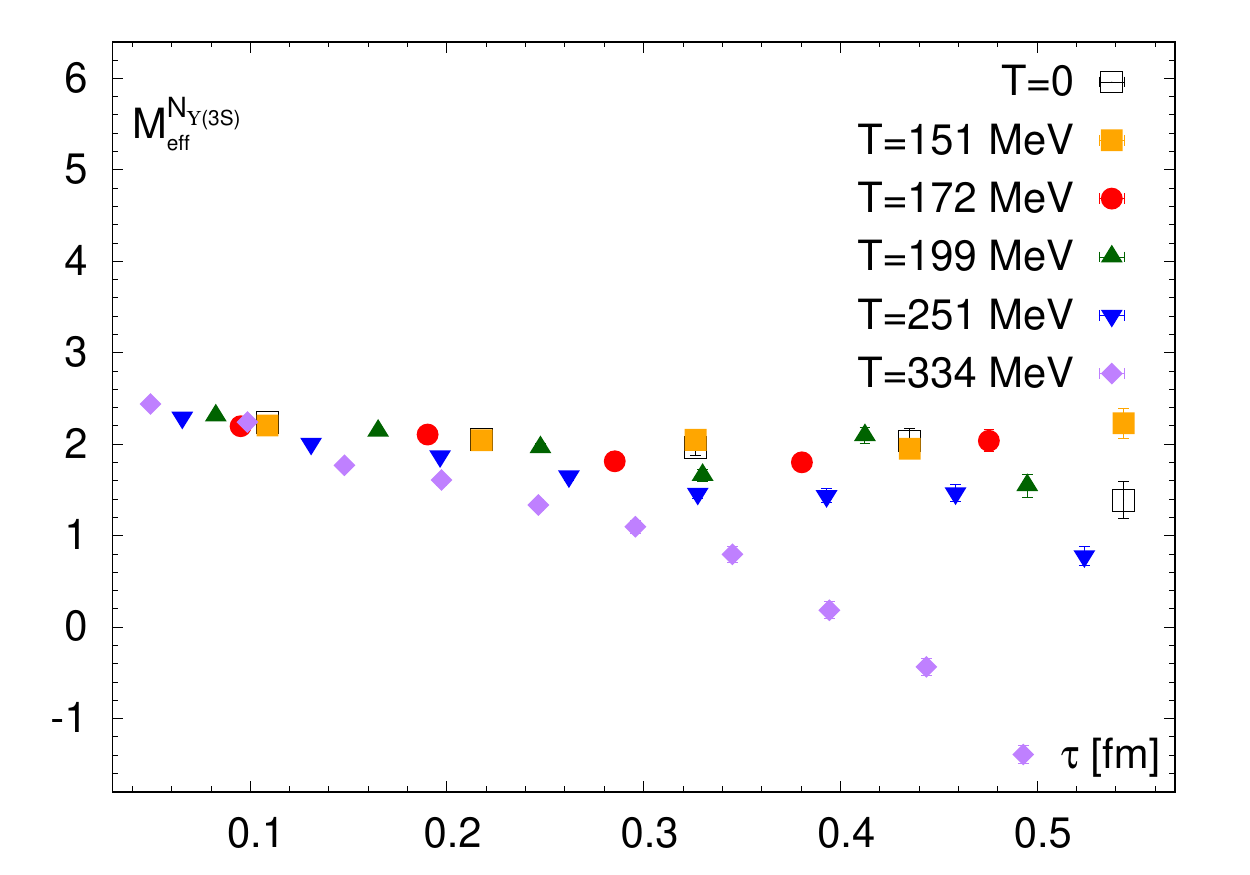}
\caption{Effective mass $M_{\mathrm{eff}}^{N_{\alpha}}$ in GeV at different temperatures for $\Upsilon(1S)$ (Top),
$\Upsilon(2S)$ (Middle) and $\Upsilon(3S)$ (Bottom) correlators. }
\label{fig:MeffN}
\end{figure}

We could also analyze the $\tau$-dependence of $N_{\alpha}(\tau,T)$ in terms of the corresponding
effective masses
\begin{equation}
  a M_{\mathrm{eff}}^{N_{\alpha}}(\tau,T) = \ln(
  \frac{N_{\alpha}(\tau,T)}{N_{\alpha}(\tau+a,T)} ) .
\end{equation}
At large $\tau$ these effective masses should reach a plateau equal to $2 E_{\alpha}$. Our
results for $M_{\mathrm{eff}}^{N_{\alpha}}$ for the different $\Upsilon(nS)$ states are shown in Fig. \ref{fig:MeffN}.
As before the effective masses have been calibrated with respect to the energy of $\eta_b$ state at $T=0$.
We see that at $T=0$ as well as at the lowest temperature the effective masses reach a plateau corresponding
to the physical mass (energy), but at higher temperatures this is not the case, in general.
For the ground state the errors are large enough so that no clear medium shift can be seen, except
at the highest temperature, $T=334$ MeV. For the $\Upsilon(2S)$ the corresponding
effective masses decrease with increasing $\tau$ for $T \ge 251$ MeV. For the $\Upsilon(3S)$
we see a significant shift in $M_{\mathrm{eff}}^{N_{\alpha}}(\tau,T)$ already for $T>191$ MeV.
The behavior of $M_{\mathrm{eff}}^{N_{\alpha}}(\tau,T)$
is qualitatively similar to the behavior of the effective masses of the correlator of optimized operators studied
in Ref. \cite{Larsen:2019zqv}. This corroborates the findings of Ref. \cite{Larsen:2019zqv} on the in-medium modifications
of the bottomonium spectral functions. For the $\Upsilon(1S)$ state our findings are also consistent with other studies
of bottomonium at non-zero temperature using NRQCD \cite{Aarts:2010ek,Aarts:2011sm,Aarts:2012ka,Aarts:2014cda,Kim:2014iga,Kim:2018yhk}.

Before concluding this section, we mention that so far we only discussed $\Upsilon(nS)$ states but very
similar results have been obtained for $\eta_b(nS)$ states as well.

\section{Comparisons between \(\mathbf{T>0}\) and \(\mathbf{T=0}\) Bethe-Salpeter
amplitudes}

\begin{figure}
\includegraphics[width=8cm]{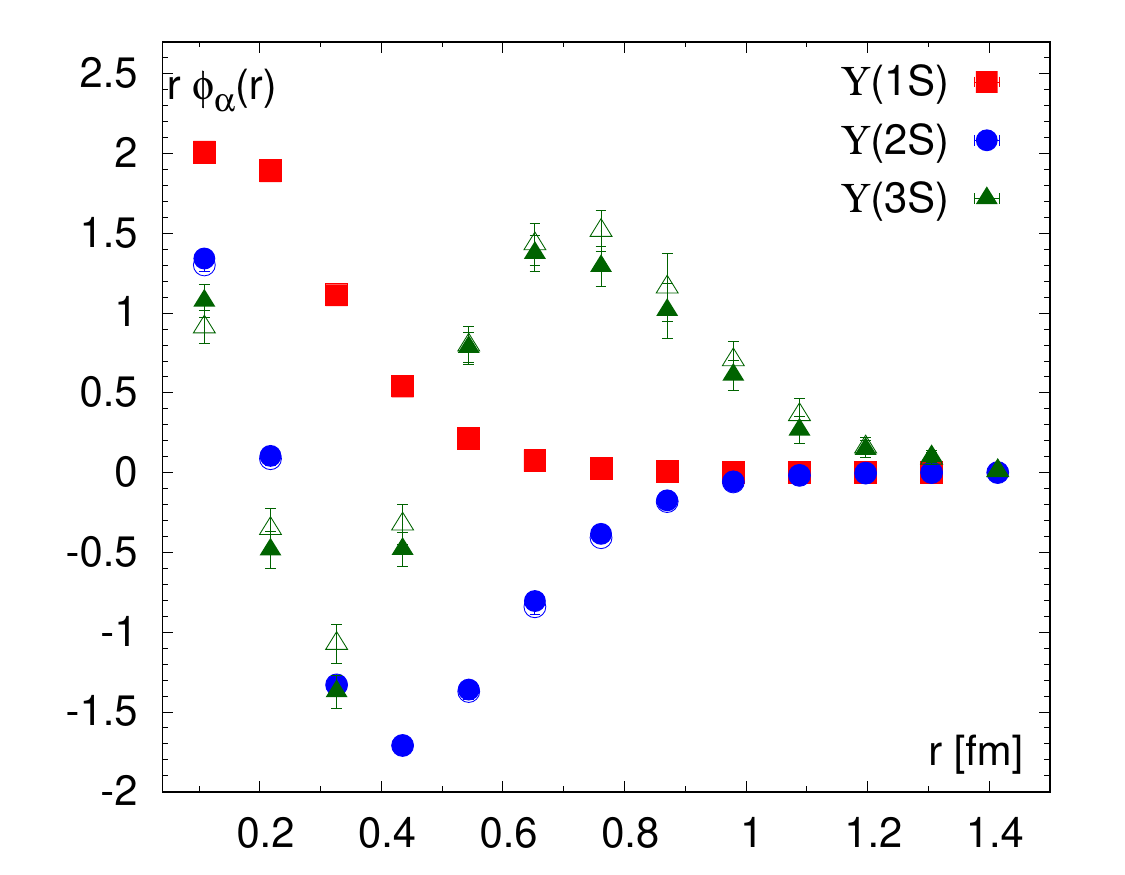}
\caption{The BS amplitudes times $r$ for the $\Upsilon(1S),~\Upsilon(2S)$ and $\Upsilon(3S)$ at $T=0$ MeV (filled symbols)
and $T=151$ MeV (open symbols) for $\tau =0.653$ fm. }
\label{fig:bs_low}
\end{figure}
If we insist on the interpretation of the correlator $\tilde C_{\alpha}^r(\tau,T)$ in terms of the wave function,
we could simply divide it by $N_{\alpha}(\tau,T)$ and study the $r$-dependence of the corresponding ratio for
sufficiently large $\tau$. At small temperatures, this ratio will have an $r$-dependence that closely follows
the $r$-dependence of the BS amplitude at $T=0$. In Fig. \ref{fig:bs_low}, we compare
$\phi_{\alpha}(\tau,T)=\tilde C_{\alpha}^r(\tau,T)/N_{\alpha}(\tau,T)$ for the lowest temperature,
$T=151$ MeV, and $\tau=0.653$ fm with the corresponding zero temperature BS amplitudes. For the $\Upsilon(1S)$ and
the $\Upsilon(2S)$, we do not see any difference between the zero temperature BS amplitude and $\phi_{\alpha}(\tau,T)$.
For the $\Upsilon(3S)$,
some difference between the zero temperature and finite temperature result for
$\phi_{\alpha}(\tau,T)$ can be seen at large
$r$, though it is not statistically significant. In any case, the $r$-dependences of $\phi_{\alpha}$
at $T=0$ and $T=151$ MeV are quite similar even for the $\Upsilon(3S)$. The lack of medium effects in
the BS amplitude for $T=151$ MeV is not surprising since at this temperature all bottomonia should
exist as well-defined states.
\begin{figure}
\includegraphics[width=8cm]{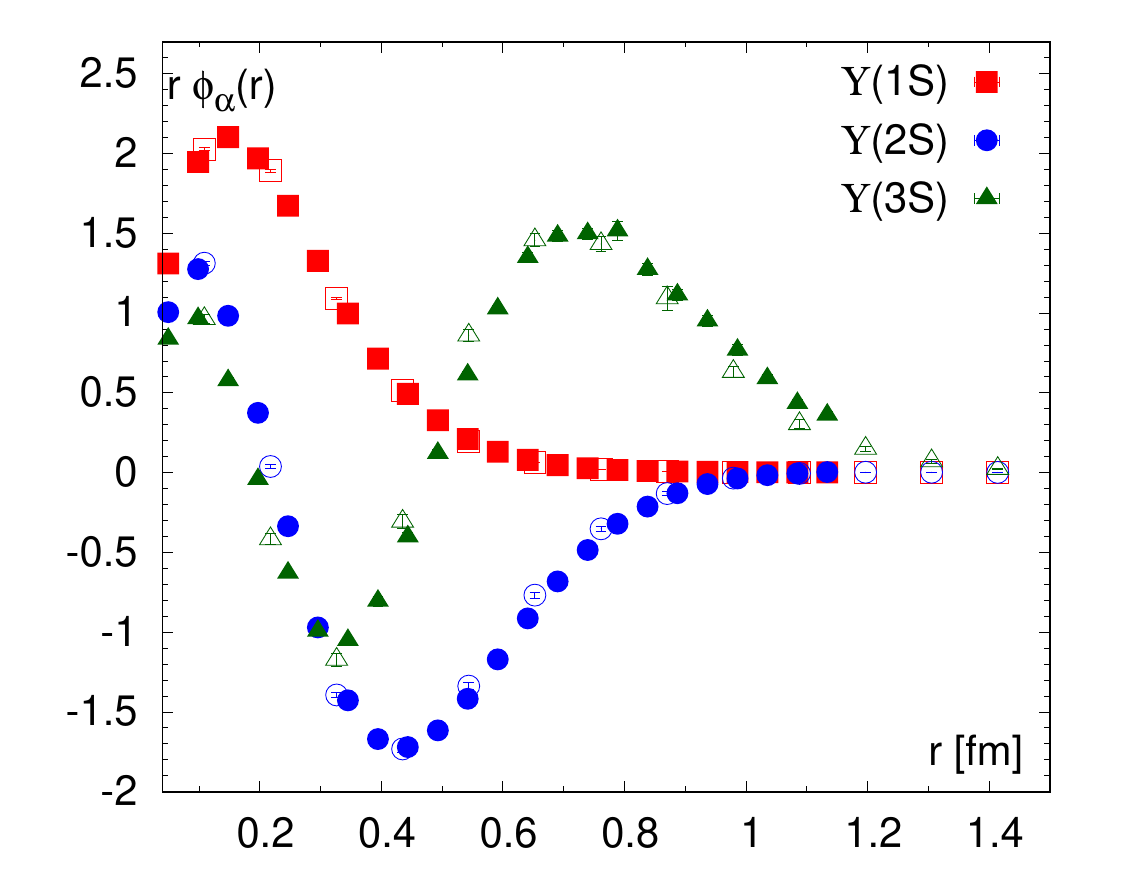}
\caption{BS amplitude times $r$  for the $\Upsilon(1S),~\Upsilon(2S)$ and $\Upsilon(3S)$ at $T=334$ MeV (filled symbols)
and $T=151$ MeV (open symbols) at $\tau \sim 0.4$~fm (see text). }
\label{fig:BS_low_high}
\end{figure}
\begin{figure}[h]
  \includegraphics[width=8cm]{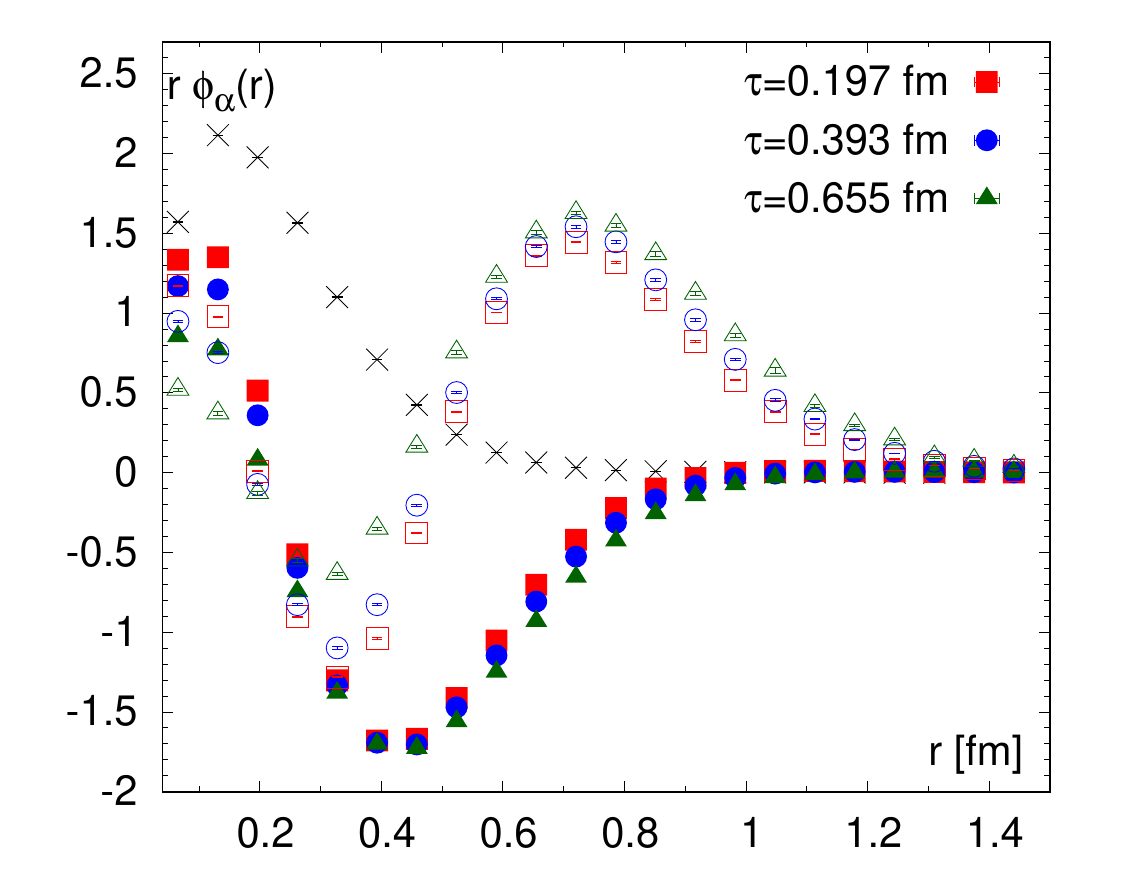}
   \caption{The BS amplitude times $r$  for the $\Upsilon(2S)$ (filled symbols) and $\Upsilon(3S)$ (open symbols) at $T=251$ MeV
   for $\tau=0.197,~0.393$ and $0.653$ fm. Also shown as crosses is the result for the $\Upsilon(1S)$.}
  \label{fig:BS_taudep}
\end{figure}
Next, we compare $\phi_{\alpha}(\tau,T)$ at the lowest and the highest temperatures for
$\tau$ around $0.4$ fm. 
Namely, we use $\tau=0.436$ fm at the lowest temperature and $\tau=0.394$ fm at the highest
temperature.
This comparison is shown in Fig.~\ref{fig:BS_low_high}, and no
temperature effect can be observed. This is presumably due to the fact that for this
$\tau$ value the contribution of the second term in Eq.~(\ref{decomp_final}) is too
small. Therefore, in Fig.~\ref{fig:BS_taudep}, we show our results for
$\phi_{\alpha}(\tau,T)$ at $T=251$ MeV and several values of $\tau$. As one can see
from the figure for the $\Upsilon(2S)$ and $\Upsilon(3S)$, there is a significant
$\tau$-dependence of $\phi_{\alpha}$. 
At small $r$, the $\tau$-dependence is mostly due to the $\tau$-dependence 
of the normalization factor $N_{norm}$ of the BS amplitude, cf. Fig. \ref{fig:Norm}, while
for larger $r$, also the shape of the BS amplitudes changes.
This suggests that the normalized
BS amplitude cannot be interpreted simply as the wave function of in-medium
$\Upsilon$ in the potential model picture. Yet, the $r$ dependence of
$\phi_{\alpha}(\tau,T)$ does not change much from one $\tau$ value to another. In
summary, the correlation $\tilde C_{\alpha}^r(\tau,T)$ shows significant temperature
dependence as one would expect based on the previous studies. However, the $r$
dependence of this correlator does not change significantly as the temperature and
$\tau$ is varied. Thus, focusing only on the $r$ dependence of $\tilde
C_{\alpha}^r(\tau,T)$  without a detailed study of its $\tau$ dependence may result
in wrong conclusions about the fate of $\Upsilon(2S)$ and $\Upsilon(3S)$ states at
high temperature. For the $\Upsilon(1S)$, there is only little dependence of
$\phi_{\alpha}$ on $\tau$, and therefore in Fig.~\ref{fig:BS_taudep}, we only show the
numerical results for $\tau=0.393$ fm. This lack of $\tau$-dependence indicates that
$\Upsilon(1S)$  can exist in the deconfined medium at $T=251$ MeV as a well defined
state with little medium modification, in agreement with the previous studies
of bottomonium at non-zero temperature based 
on NRQCD \cite{Aarts:2010ek,Aarts:2011sm,Aarts:2012ka,Aarts:2014cda,Kim:2014iga,Kim:2018yhk}.

The lack of temperature dependence of the normalized BS amplitude at $T>0$ at
$\tau \simeq 0.4$ fm demonstrated in Fig. \ref{fig:BS_low_high} has an interesting
consequence. It means that $\phi_{\alpha}(r,T)$ can be used as a proxy for the $T=0$
BS amplitude at zero temperature. Since the two temperatures shown in Fig.
\ref{fig:BS_low_high} correspond to two different lattice spacings this result also
implies that the lattice spacing dependence of the BS amplitude is small.  Therefore,
the comparison of the wave function obtained from potential model and BS amplitude
obtained on the lattice with $a=0.1088$ fm in Section II seems justified.

\section{Conclusions}

Using lattice NRQCD in this paper, we studied the correlation functions, $\tilde
C_{\alpha}^r$, between operators optimized to have good overlaps with the of
$\Upsilon(1S)$, $\Upsilon(2S)$, and $\Upsilon(3S)$ vacuum wave functions and simple
spatially non-local bottomonium operators, where the bottom quark and anti-quark are
separated by distance $r$. This correlator has been calculated at zero as well as at
non-zero temperature. At zero temperature, $\tilde C_{\alpha}^r$ can be interpreted in
terms of the Bethe-Salpeter amplitude. We have found that the $r$-dependence of the
Bethe-Salpeter amplitude closely resembles the corresponding potential model based
bottomonium wave function. Moreover, by choosing the bottom quark mass 
used in the Schr\"odinger equation to be approximately $5.5$~GeV we estimated the heavy quark antiquark
potential from Bethe-Salpeter amplitudes and found agreement with the static quark
potential calculated on the lattice. These findings support the potential model for the
bottomonium in vacuum.

We studied the temperature and Euclidean time dependence of $\tilde C_{\alpha}^r$ in
terms of effective masses. For $\Upsilon(1S)$, we see only very small temperature and
Euclidean time dependence of the corresponding effective masses, except at the
highest temperature of $334$~MeV. For $\Upsilon(2S)$ and especially for
$\Upsilon(3S)$ significant dependence on the Euclidean time were observed, making it
difficult to draw parallels between Bethe-Salpeter amplitudes and potential model
based in-medium wave functions.  Since the $r$-dependence changes very little with
varying Euclidean time and temperature, focusing solely on the $r$-dependence of
$\tilde C_{\alpha}^r$ at a fixed $\tau$ might lead to misleading conclusions
regarding existence of well-defined $\Upsilon(2S)$ and $\Upsilon(3S)$ in medium. On
the other hand, we found that the behavior of the effective masses is similar to the
one previously studied by us using correlators of optimized bottomonium
operators~\cite{Larsen:2019zqv}, supporting the picture of thermal broadening of
bottomonium states.

\section*{Acknowledgments}
This material is based upon work supported by: (i) The U.S. Department of Energy,
Office of Science, Office of Nuclear Physics and High Energy Physics through the
Contract No. DE-SC0012704; (ii) The U.S. Department of Energy, Office of Science,
Office of Nuclear Physics and Office of Advanced Scientific Computing Research within
the framework of Scientific Discovery through Advance Computing (ScIDAC) award
Computing the Properties of Matter with Leadership Computing Resources. (iii) S.
Meinel acknowledges support by the U.S. Department of Energy, Office of Science,
Office of High Energy Physics under Award Number DE-SC0009913. (iv) Computations for
this work were carried out in part on facilities of the USQCD Collaboration, which
are funded by the Office of Science of the U.S. Department of Energy. (v) This
research used awards of computer time provided by the INCITE and ALCC programs at Oak
Ridge Leadership Computing Facility, a DOE Office of Science User Facility operated
under Contract No. DE-AC05- 00OR22725.
\bibliography{ref}
\end{document}